\documentstyle[twocolumn]{mn}
\title[Collisional versus Collisionless Matter]
{Collisional versus Collisionless Matter: a
One-dimensional Analysis of Gravitational Clustering}
\author[Claudio Gheller, Lauro Moscardini
and Ornella Pantano]{Claudio Gheller$^{1}$,
Lauro Moscardini$^{2}$
and Ornella Pantano$^{3}$ \\
$^{1}$ Scuola Internazionale Superiore di Studi Avanzati, SISSA,
via Beirut 2--4, I--34013 Trieste, Italy\\
$^{2}$ Dipartimento di Astronomia, Universit\`{a} di Padova, vicolo
dell'Osservatorio 5, I--35122 Padova, Italy\\
$^{3}$ Dipartimento di Fisica {\em Galileo Galilei}, Universit\`{a} di
Padova, via Marzolo 8, I--35131 Padova, Italy\\}

\begin{document}

\maketitle

\begin{abstract}
We present the results of a series of one-dimensional N-body and hydrodynamical
simulations which have been used for testing the different clustering
properties of baryonic and dark matter in an expanding background. Initial
Gaussian random density perturbations with a power-law spectrum $P(k) \propto
k^n$ are assumed. We analyse the distribution of density fluctuations and
thermodynamical quantities for different spectral indices $n$ and discuss the
statistical properties of clustering in the corresponding simulations. At large
scales the final distribution of the two components is very similar while at
small scales the dark matter presents a lumpiness which is not found in the
baryonic matter. The amplitude of density fluctuations in each component
depends on the spectral index $n$ and only for $n=-1$ the amplitude of baryonic
density fluctuations is larger than that in the dark component. This result is
also confirmed by the behaviour of the bias factor, defined as the ratio
between the r.m.s  of baryonic and dark matter fluctuations at different
scales: while for $n=1,\ 3$ it is always less than unity except at very large
scales where it tends to one, for $n=-1$ it is above 1.4 at all scales. All
simulations show also that there is not an exact correspondence between the
positions of largest peaks in dark and baryonic components, as confirmed by a
cross-correlation analysis. The final temperatures depend on the initial
spectral index: the highest values are obtained for $n=-1$ and are in proximity
of high density regions.
\end{abstract}


\section{Introduction}
In recent years a considerable improvement in the study of large-scale
structure has been obtained with the development of three-dimensional codes
able to follow simultaneously the evolution of dark and baryonic matter. It is
important to include the gas in N-body simulations for at least two reasons:
first we must estimate the large amount of gas that is observed in clusters of
galaxies through X-ray emission; second, we must remove any ambiguity in the
identification of galaxies. In the numerical simulations, the hydrodynamical
equations have been solved either using mesh based methods
\cite{crv89,c92,rokc93,bnsco94} or particle methods like the smoothed particle
hydrodynamics \cite{hk89,e90,kg91}. The latter, being Lagrangian methods, have
the advantage of covering a larger range of clustering scales, but they are not
very reliable in low density regions for the calculation of thermodynamical
quantities. Eulerian methods, on the contrary, give a better estimate of
thermodynamical quantities, but are constrained by the limited grid resolution
in their capability of following the collapse of high density regions. A large
dynamical range can be attained only by increasing the number of grid-points,
but this is obviously limited by the available computer memory. At present the
largest mesh used in Eulerian calculations has  $256^3$ grid-points. The
analysis of the evolution at different cosmological scales is obtained by
changing the physical normalization of the box size, and sometimes the results
obtained at a given scale are used for setting up initial and boundary
conditions  at another scale (e.g. Cen \& Ostriker 1992).

Since three-dimensional calculations pose severe limits on the number of
grid-points used in Eulerian methods, there has been in recent years an attempt
to make up for the lack of resolution with the use of more accurate algorithms.
One is looking for numerical methods able to well describe both low density
regions and strong gradients in the fluid flow. Very interesting is, for
example, the possibility of introducing mesh refinement schemes in grid-based
codes which, associated with shock capturing methods, would considerably
improve the resolution in high density regions \cite{anc94}.

The limited resolution reachable in three-dimensional codes makes sometimes
difficult to separate, in the discussion of results, those effects which are
only numerical from those which contains real physical information. Analysis
like the one of Kang et al. (1994) are very useful for discriminating between
different methods in three dimensions, but, because of the limitation in
computer memory affecting all of the methods, a simultaneous one-dimensional
analysis of the same phenomena appears quite convenient. Working only in a
single dimension makes possible to span a large dynamical range and to follow
structure formation at different scales. Approximate analytic solutions and
their behaviour in highly non-linear regimes can also be better tested in
one-dimensional calculations (e.g. Williams et al. 1991). Recently, Quilis,
Ib\'anez \& Saez (1994) have studied the applicability of modern
high-resolution shock-capturing methods to the study of cosmological structures
in presence of pressure forces. However, they do not follow at the same time
the dynamics of the dark matter and so they are not able to compare the
clustering properties of the two different components. In our work, instead, we
study the evolution of cosmological one-dimensional perturbations when both
baryonic and dark matter are considered.

A detailed study of the evolution of gas and collisionless matter in a single
pancake was presented by Bond et al. (1984) and Shapiro \& Struck-Marcell
(1985). In both of these works one-dimensional calculations of the coupled
evolution of baryonic and dark component were performed including the effects
of radiative and Compton cooling terms and thermal conductivity of the gas. The
main difference between these two analyses is in the choice of the numerical
algorithm used for solving the hydrodynamical equations which was Lagrangian in
one case \cite{ss85} and Eulerian in the other one \cite{bcsw84}. A similar
analysis has been recently done by Thoul \& Weinberg (1994), for a spherical
configuration. They also have used a Lagrangian approach which has the
advantage of giving a higher resolution where needed without having, in
one-dimension, the same problems of grid distortion present in more dimensions.
All of the previous one-dimensional works following the coupled evolution of
dark and baryonic matter consider isolated perturbations in the expanding
universe and start their simulations when hydrodynamical effects begin to
become important in the evolution. The information concerning the formation
epoch and mass of these isolated structures is derived from N-body calculations
assuming some model of large-scale structure formation.

In this work we do not limit our attention to isolated perturbations but we
want to explore the differences in the statistical properties of the
one-dimensional clustering of baryonic and dark matter (hereafter BM and DM,
respectively) components in an expanding background. We analyse the effect of
pressure forces and adiabatic heating in the dynamics of the gas and we do not
include, at this stage, cooling terms in our computations. Our aim is to
understand first the evolution of one-dimensional structures only in the
presence of ``compressible'' effects. In addition, the inclusion of cooling
processes requires the specification of characteristic time and length scales
in the problem and in one-dimensional cosmological calculations it is not
obvious how one can choose these quantities in connection with theory and
observations. Therefore, we have preferred to make simulations which are
scale-free; the absence of features in the primordial power-spectrum together
with the higher spatial resolution implied by the use of one-dimensional
simulations would permit to better discriminate between numerical artifacts and
physical effects in the results. At the end, physical information on the
computed quantities can easily be extracted by fixing the normalization
quantities, like the box size and the final time.

The plan of the paper is as follows. In Section 2 we introduce the dynamical
equations both for the hydrodynamical variables and for collisionless matter.
The numerical methods used for the solution of these equations are also
presented together with some numerical tests. The results obtained in numerical
simulations of cosmological one-dimensional structures are shown in Section 3.
Discussion and conclusions are drawn in Section 4.

\section{Dynamical Equations}

In this section we present the dynamical equations that we use for evolving the
BM and DM components in the case of one-dimensional perturbations. We assume a
flat universe ($\Omega=1$) in which the BM component accounts for $10\%$ of the
total mass ($\Omega_{_{BM}}=0.1$) and the rest is in the form of DM
($\Omega_{_{DM}}=0.9$). In the present analysis we neglect the effects of
radiative and Compton coolings and any possible external heating. The DM
component is approximated as a pressureless fluid and the BM component as a
perfect fluid.

\subsection{Hydrodynamical equations}

We introduce the following set of dimensionless variables:
\begin{eqnarray}
\tilde t &=& {t \over t_0}, \\
\tilde x &=& {x \over x_0}, \\
\tilde v &=& v {t_0\over x_0}, \\
\tilde {\varrho} &=& {a^3 \varrho \over \varrho_0}, \\
\tilde {\epsilon} &=& a^3 \epsilon {t_0^2\over \varrho_0 x_0^2 }, \\
\tilde {\phi} &=& \phi {t_0^2\over x_0^2 },
\end{eqnarray}
where $x$ is the comoving coordinate, $t$ is the time, $a= \tilde t^{~2/3}$ is
the cosmological expansion factor, $\phi$ is the peculiar gravitational
potential, $\varrho$ and $\epsilon$ are the matter and internal energy
densities, respectively. We have three basic units of normalization: the final
time $t_0$, the comoving cell size $x_0$ and the mean baryon density
$\varrho_0$ at time $t_0$. For a cosmological application we can take $t_0$ and
$\varrho_0$ equal to the present age and density of the Universe, respectively.

The evolution of the collisional component is described by the following set of
hydrodynamical equations:
\begin{equation}
{\partial\tilde \varrho\over \partial \tilde t}+{1\over
a}{\partial\over\partial \tilde x}[\tilde \varrho \tilde v] = 0\ ,
\end{equation}

\begin{equation}
{\partial\tilde \varrho \tilde v\over\partial \tilde t}+{1\over
 a}{\partial\over\partial \tilde x}
[\tilde \varrho \tilde v^2+ \tilde p] =
-{\dot a\over a}\tilde \varrho \tilde v-{1\over
 a}\tilde \varrho{\partial\tilde \phi\over\partial\tilde x}\ ,
\end{equation}

\begin{equation}
{\partial \tilde \epsilon \over \partial \tilde t}+{1\over a}
{\partial \over \partial \tilde x}[\tilde \epsilon \tilde v] =
-2{\dot a\over a}\tilde \epsilon-{1\over a} \tilde p{\partial \tilde v
\over\partial \tilde x}\ ,
\end{equation}
where $\tilde p$ is the comoving pressure.

We use the equation of state of an ideal gas with adiabatic index $\gamma =
5/3$, so that
\begin{equation}
\tilde p = (\gamma-1)\tilde \epsilon \ .
\end{equation}

The physical temperature $T$ is related to the previous normalized quantities
by
\begin{equation}
T = 2 {m_p\over K_B} ({x_0 \over t_0})^2 {\tilde \epsilon \over \tilde
\rho} \ ,
\end{equation}
where $K_B$ is Boltzman's constant and $m_p$ the mass of the proton.

\subsection{Collisionless matter}

In the dynamical equations for the collisionless particles we use as time
integration variable the scale factor $a$:
\begin{equation}
{d \tilde x\over da}=\tilde u\ ,
\end{equation}
\begin{equation}
{d \tilde u\over d a}+{3\over 2a}\tilde u = - {9\over 4a}{\partial\tilde
\phi\over\partial \tilde x}\ .
\end{equation}

The peculiar gravitational potential due to both the BM and DM component is
computed by solving the Poisson equation:
\begin{equation}
{\partial^2\tilde \phi\over\partial \tilde x^2} = {K\over
 a}(\tilde\varrho_{_{BM}}+\tilde\varrho_{_{DM}}-1)\ ,
\end{equation}
where $K = 4\pi G \varrho_0 t_0^2$ and $G$ is the Newton's constant.

\subsection{Numerical Methods}

The hydrodynamical equations have been integrated using the Flux Corrected
Transport (FCT) method \cite{bb73,z79}, a hybrid shock capturing method. In
this method two difference schemes are blended together: a second order
Lax-Wendroff scheme is used in regions of smooth flow, while a first order Lax
scheme is used near discontinuities. Then second order accuracy is ensured
everywhere except near flow jumps where the  dissipation introduced by the low
order scheme guarantees monotonicity in the behaviour of flow variables. The
blending between the two schemes is controlled by a monotonicity constraint
that leads to the sharpest possible discontinuity profiles. However, if the
grid is not sufficiently fine, very strong gradients can be represented as a
sequence of discontinuous jumps. This effect usually does not hinder the
convergence of the method and can be reduced or completely avoided refining the
grid \cite{wc84}.

For the integration of the collisionless matter we have used a particle-mesh
code \cite{he81}. We use the same mesh as in the hydrodynamical part of the
code, while the number of particles is a multiple of the number of mesh points
in order to have a good resolution also in low density regions. The
interpolation used for computing the mass density and the forces acting on each
particle is obtained by a TSC scheme (e.g. Hockney \& Eastwood 1981), which
ensures a good accuracy in the estimates of previous quantities without leading
to an excessive slowing down of the computation. For the time integration we
use a second order leap-frog method so that the accuracy in the N-body is
comparable to that attained in the hydrodynamical part of the code. The
peculiar gravitational potential is computed using the standard fast Fourier
transform technique.

The accuracy of the code has been tested against known analytical solutions.
For the hydrodynamical part we performed the shock tube test comparing the
analytical and numerical solutions. We found a very good agreement in smooth
regions, while the shock is numerically diffused over about five grid-points.
The evolution of an initial sinusoidal perturbation has been instead used for
testing the performance of the N-body code: in this case the numerical solution
has been compared with the results obtained using the Zel'dovich approximation
\cite{z70} which provides the exact one-dimensional solution up to shell
crossing. In Figure 1 we compare, just before shell crossing, the resulting DM
distributions for the Zel'dovich and N-body solutions for two  different
choices of the number of grid-points $N_g$ ($N_g=2^{10}$ and $N_g=2^{15}$). The
number of particles is instead kept fixed and set equal to $2^{16}$. We notice
that already with the smallest number of grid-points the N-body solution is in
very good agreement with the Zel'dovich result. The difference in the height of
the peak in the two cases is only due to the different numerical resolution
implied by the adopted grids.

We used a sinusoidal perturbation also for testing the coupled evolution of
dark and baryonic components and to see how far the code is able to follow the
collapse of a baryonic fluctuation. In Figure 2 we show the density profile of
both components after the shell crossing time, obtained using a grid of
$2^{15}$ points. We can see that fluctuations in the baryonic matter as larger
as $10^2$ can be easily resolved with this resolution. Some typical problems of
the FCT method start to appear after the formation of strong shocks which are
not well resolved using the previous grid, but they do not destroy the solution
and a local spatial refinement would permit to continue the evolution. Since
situations as this one are never found in the following cosmological
simulations, we believe that a grid of $2^{15}$ is sufficient for our purposes.

\section{Cosmological One-dimensional Structures}

In this section we present the results of the evolution of one-dimensional
perturbations in an expanding background. We analyse the distribution of BM and
DM, paying particular attention to the clustering properties of the two
components and to the effects due to their coupling.

\subsection{Numerical Simulations}

We consider a grid of comoving length $L$, whose physical size grows according
to the expansion parameter $a$. We subdivide our grid into $2^{15}$ nodes and
adopt periodic boundary conditions. The number of particles used in the N-body
part of the code is $2^{16}$.

The initial conditions for a given model are determined as follows. The DM
component is perturbed  according to a Gaussian random density field
characterized by a power-law spectrum of the general form
\begin{equation}
P(k) = A k^n {\rm exp} (-k^2 R_f^2) \ ,
\end{equation}
where $A$ is a normalization constant and $n$ the spectral index. The short
wavelength cut-off at the scale $R_f=2$ grid-points ensures that the results
are not affected by the sampling of modes whose size is close to that of the
resolution of the simulation.

The normalization constant $A$ is fixed requiring that the one-dimensional
variance $\sigma^2$ at the final time $t_0$ (corresponding to the expansion
factor $a=1$) for linearly evolved perturbations is equal to unity. In
computing the variance we assume a filtering radius $R_{\star}=10^{-2} L$ so
that
\begin{equation}
\sigma^2(R_{\star}) = {1\over \pi}\int^{\infty}_{0} P(k)
W^2(k,R_{\star})\ dk\ .
\end{equation}
The function $W(k,R_{\star})$ is a one-dimensional top-hat filter:
\begin{equation}
W(k,R_{\star}) = {\sin (kR_{\star})\over kR_{\star}}\ .
\end{equation}

Initially the DM particles are evolved according to the Zel'dovich (1970)
algorithm which is known to give the exact solution up to the shell crossing.
We start our  numerical computation when one of the following requirements is
satisfied: i) the largest density fluctuation is equal to unity; ii) the first
shell crossing has occurred: this typically corresponds to epochs with $a <
10^{-3}$ for all models.

The numerical evolution of BM starts when $a=10^{-3}$, i.e. just after the
recombination epoch. Before this time we assume that the distribution of BM is
constant, as it would be in comparison with DM distribution, because of the
coupling between radiation and baryonic matter before hydrogen recombination.
The internal energy is fixed assuming that the cooling of the gas is due only
to adiabatic expansion between the epoch of recombination and the initial time
of the simulations. In the subsequent evolution the perturbation in the
baryonic matter is induced by gravitational coupling with the DM component.

As already pointed out, our simulations are completely scale-free: the final
amplitude of the fluctuations depend only on the scale $R_\star$ relative to
which the initial normalization is fixed. The final thermodynamical quantities
are completely determined once the amplitude of initial perturbations and the
initial value of $\tilde \epsilon$ are given. For example, if we think of
$R_\star$ as corresponding of $8~h^{-1}$ ($h$ is the Hubble constant in units
of 100 km s$^{-1}$ Mpc$^{-1}$), where observationally the variance is found to
be unity, the whole box would correspond to $L=800~h^{-1}$ Mpc (we use
$R_\star=L/100$) and then we would be looking the evolution at very large
scales. In order to study the evolution on smaller scales but fixing the
normalization of the initial spectrum at the same physical scale, we should use
a larger value for $R_\star$ in units of $L$ and, consequently, we should start
with larger density fluctuations. These simulations would produce larger final
temperatures on these smaller scales as we expect.

\subsection{Results}

We consider three different models with primordial spectral index $n=-1,\ 1, \
3$: in three dimensions, these would correspond to $n=-3,\ -1, \ 1$ (the
so-called Harrison-Zel'dovich spectrum) respectively, covering in this way the
range of values usually adopted for cosmological models. For each model we run
three simulations with different realizations of the initial conditions in
order to obtain more accurate estimates of the relevant quantities. Here we
consider various statistical tests, such as the distribution of density
perturbations, peaks and thermodynamical quantities, their correlations and the
density power-spectrum. The results referring to the DM component have been
previously smoothed by a Gaussian filter with a radius of ten grid-points,
roughly corresponding to twice the numerical spreading of the shock front in
the hydrodynamical calculations.

\subsubsection{Distribution of Density Perturbations}

In Figures 3a and 3b we show the behaviour of $\delta_{_{BM}}$ and
$\delta_{_{DM}}$ in a realization for each model at two different epochs:
$a=0.5$ and the final one $a=1$, respectively. The results refer to about a
tenth of the whole grid. Different runs of the same model show a very similar
qualitative behaviour. The density fluctuations are computed with respect to
the mean value of the corresponding component. As expected, increasing the
spectral index $n$, we have less power on large scales and, consequently, peaks
in density contrast are more frequent, but they have a smaller density
contrast, while underdense regions are less extended. We notice that only for
$n=-1$ peaks in BM  are higher than those in DM  and their contribution to the
local gravitational field starts to be comparable to that of DM.

For the $n=3$ case there is a correspondence between the distribution of DM and
BM although there are always substructures in the collisionless component which
are not present in the other one (see also the following discussion on relative
bias, power-spectrum and number of peaks). This behaviour is enhanced as we
decrease the spectral index: BM appears more and more clumped while DM stays
more spread with small substructures. This is particularly evident in the
structures appearing in the model with $n=-1$, where collisionless matter is
spread over a region which is several times larger than that occupied by the
peak in BM.  It is interesting to notice also that there is not an exact
correspondence between the positions of the largest BM and DM peaks.

Comparing the figures at two different times we see the tendency to the merging
of substructures which produces higher peaks and larger voids. This behaviour
is again more evident in simulations with more power on large scales. For
example, in the panels referring to the $n=-1$ simulation, it is possible to
note that the two central peaks in the BM component at $a=0.5$ quickly merge
forming a unique high-density region at $a=1$. This phenomenon is less
pronounced for DM, which at $a=1$ continues to be divided into small
substructures.

A quantitative estimate of the spatial correspondence between the BM and DM
density distributions can be obtained studying the cross-correlation
coefficient (see e.g. Coles, Melott \& Shandarin 1993)
\begin{equation}
S(R)={\langle\delta_{_{BM}}(R)~\delta_{_{DM}}(R)\rangle\over
\sigma_{_{BM}}(R)~\sigma_{_{DM}}(R)}\ ,
\end{equation}
where $\delta_{_{BM}}$ and $\delta_{_{DM}}$ are respectively the density
contrasts in the BM and DM components, smoothed with a Gaussian window of
radius $R$ and $\sigma_i=\langle\delta_i^2\rangle^{1/2}$ are the corresponding
r.m.s.. The mean is computed over all grid-points. Definition (18) implies
$\mid S\mid\leq 1$. The limit $S=+1$ corresponds to
$\delta_{_{BM}}=C\delta_{_{DM}}$, where $C$ is a constant: in this case there
is a perfect agreement in the positions of the structures in the two
components. In Figure 4, the cross-correlation coefficient $S$ is plotted as a
function of the Gaussian filtering radius $R$ for the three models when $a=0.5$
and $a=1$. The error bars, shown for clarity only at the final time, represent
the scatter r.m.s. between the three different realizations of each model. At
earlier times these errors are found to be always smaller. As expected,
decreasing the filtering radius and/or allowing for a longer evolution, the
differences between the two components increase and, consequently, the
coefficient $S$ decreases. This is particularly true for the model with $n=-1$,
where, for small $R$ (up to $L$/400), the cross-correlation is even less than
0.4, denoting the small correspondence between BM and DM distributions which
was already clear in Figure 3. On the contrary, a value of $S$ close to unity
is obtained in the simulations with $n=3$, even when a small filtering radius
is adopted.

The study of high-density regions in BM is particularly interesting because
they are expected to be related to the positions where galaxy formation occurs.
We define as peaks all the grid-points which are local maxima of the density
fluctuation field. In Figure 5 we show, at two different times, the number of
peaks $N_{pk}$ having an height greater than a given threshold $\delta$, both
for BM and DM components. The comparison between the models with different
spectral indices shows once again a different behaviour. Only for the model
with $n=-1$, the number of high peaks in BM is larger than that of DM at both
considered times. The latter model is also the one which shows the largest time
evolution in the number of peaks. In the other cases, $N_{pk}$ does not changes
significantly and the peak number in the BM component is always less than that
in DM one.

\subsubsection{Thermodynamical quantities}

In Figure 6 we show the behaviour  of the baryonic quantities $\tilde
\varrho_{_{BM}}$, $\tilde p$ and $T$ for the case $n=-1$ at $a=1$ (the results
refer to the same realization shown in the upper panel of Figure 3b but in this
case the whole grid is displayed). This is the model in which the highest
values in thermodynamical quantities are obtained, thus some effects, which are
also present in the other  models, are better visible. The baryonic density and
the pressure are given in normalized units, while the temperature is shown in
Kelvin degrees. In absence of heating from external sources and cooling
processes, we notice a strong correspondence among peaks in density, pressure
and temperature. However, in correspondence of the largest density fluctuations
we observe a double peak on the top of the temperature profile which shows, as
expected, that heating mainly occurs in the regions in which the velocity
gradient is large.

The temperature distribution in the different models is better illustrated by
the cumulative volume filling factor and mass fraction $F$ as a function of the
temperature $T$. Figure 7 shows the behaviour of these quantities at $a=1$ as
obtained by summing over the contribution of all three realizations for each
model. We notice that the fraction of matter at high temperature (e.g. $ T \ge
10^4$ K) increases as we decrease the spectral index: it is approximately 15\%
and  2\% for $n=-1$ and $n=1$, respectively, while no grid-point has such a
high temperature in the $n=3$ model (in this case the highest temperature
obtained is about 25 K).  At the same time, most of the volume is at low
temperatures: this is particularly evident in the case $n=-1$, where, due to
the dominance of empty regions, more than 90\% of the volume has $T \le
10^{-1}$ K. On the contrary, the two cumulative functions are more similar in
the model with $n=3$, where structures are smaller and more uniformly
distributed. These results suggest that, in absence of other dissipative or
heating effects (viscosity, thermoconduction, starbursts, etc.), it is
difficult to obtain from one-dimensional cosmological large-scale perturbations
temperatures as high as those seen in X-ray observations.

Figure 8 shows contour plots for the number of zones characterized by a given
temperature and baryonic density. The additional information which we can
extract from these graphs is the correlation between these two quantities in
all the three models. This correlation is particularly strong in the case
$n=-1$ and refers to the large empty regions characterizing this model. On the
contrary, more scattered distributions are presented in the models with $n=1$
and $n=3$.

\subsubsection{Power-spectrum and bias}

In order to study the time evolution of clustering at different scales, we have
computed the power-spectrum $P(k)$. If the density fluctuations remain linear
at large scales, $P(k)$ is expected to grow as $a^2$. In Figure 9 we show the
behaviour of the power-spectrum both for DM and BM for the three models at
different times. The units of the wavenumbers $k$  are such that $k=1$
corresponds to the fundamental wavelength of the computational grid.
Considering the models with $n=3$ and $n=1$, we see that at small wavenumbers
(i.e. at large scales) the growth of the perturbations in the DM component is
in good agreement with linear theory. The situation is different for the model
with $n=-1$ where the growth of perturbations is non-linear even at these
scales. At larger wavenumbers (i.e. at small scales) deviations from linear
theory are found, as expected. In particular, we notice that there is a faster
evolution for the models with $n=-1$ and $n=1$: the non-linear growth produces
the coupling of high- and low-$k$ modes. In the model with $n=3$, where
initially the large-scale power is small, this effect does not appear. Then,
our simulations, even if starting with a power-law spectrum, seems to exclude
the possibility of having self-similar evolution when large scale power is
present. Only the model with $n=3$ presents a self-similar evolution in the DM
at least for $\log k \le 2.4$.

The behaviour of the BM is very different. At the beginning, due to the
assumption of uniform distribution on all scales, the power-spectrum is zero.
At early times all models show a very slow growth produced by the coupling with
the DM fluctuations (the curves for the model $n=-1$ are not visible because
$P(k)$ is too low). Only at late epochs ($a \approx 0.5$), the fluctuations in
the BM grow considerably and quickly reach those in the DM. At the final time
$a=1$ the power-spectra in the two components are different at small scales,
where the growth of gas perturbations is delayed by the pressure: this is
particularly true in the models with $n=-1$ and $n=1$, for which higher values
of the pressure are reached.

An interesting way to measure the relative growth of the perturbations is the
bias factor $b$, defined as
\begin{equation}
b(R)= { \sigma_{_{BM}}(R) \over {\sigma_{_{DM}}(R)}}\,,
\end{equation}
where $\sigma_i^2(R)$  is the variance of the density fluctuation field in the
$i$-component, smoothed using a Gaussian window of radius $R$. Figure 10 shows
the behaviour of $b$ for the three models at $a=0.5$ and $a=1$. The scatter
r.m.s. between the different realizations of each model is also plotted only at
the final time. The models with $n=1$ and $n=3$ behave in a similar way: the DM
appears more clustered than the BM at small scales, while at larger scales they
tend to have the same clustering properties. As expected, the pressure in the
BM prevents shell-crossing and a collapse comparable to that occurred in the
DM. At larger scales where dissipation is negligible, the two distributions are
similar. In the case $n=-1$ instead, the BM at the final time $a=1$ has a
larger variance than the DM component at all scales; this is in agreement with
the density distribution shown in Figure 3b. The motivation for the different
behaviour of this model is the large peculiar velocities of the DM component.
In fact, the absence of dissipative effects delays the formation of more
compact structures. In the BM, instead, higher temperatures and pressures are
reached, but they are not sufficient to overcome the gravitational force and
avoid the collapse. In addition the conversion of the kinetic into internal
energy allows the formation of more compact baryonic structures.

{}From Figure 10 it is possible to extract also information about the time
evolution of the bias factor. Again we can note a different behaviour between
the models. In the case $n=-1$, the evolution of $b$ from $a=0.5$ to $a=1$ is
rapid at all scales: $b$ changes by a factor $\approx 1.5$. On the contrary,
the increase of $b$ in the same time interval is smaller for $n=1$ and $n=3$
(approximately 15\% and 5\%, respectively) but the final value is always below
unity.

\section{Discussion and Conclusions}

In this paper we have studied the evolution of one-dimensional perturbations in
a medium composed by BM and DM. Initial Gaussian perturbations with a power-law
spectrum are assumed in the dominant DM  component and then the perturbations
in the BM  component, initially uniformly distributed, are induced through
gravitational coupling. We observe that BM does not fall immediately  into the
potential wells created by the distribution of the collisionless component, but
when this happens, the amplitude of density perturbations in the gas becomes
quickly comparable to that of the DM. At large scales the final distribution of
the two components, as shown by the cross-correlation coefficient and by the
bias factor, is very similar. The main differences are present at small scales
where the DM  component presents a lumpiness which is not found in the BM. A
quite important feature shown by all simulations is the non-exact
correspondence between largest DM and BM peaks.

In our simulations peaks in the DM  are always higher than those in BM except
for $n=-1$, i.e. for the spectrum with more power on large scales. In this case
the variance in baryon density distribution is larger than that of DM  at all
scales. Moreover, in the highest baryon peaks the self-gravity of baryons is
comparable to the gravitational field due to the collisionless component and
this can have an important effect on the evolution of the latter. In general,
many peaks in the DM component are associated to a single peak in the BM. These
are usually embedded in the corresponding baryonic peak which stays more spread
in space because of the effect of pressure working against the collapse. For
the highest BM peaks present in the case $n=-1$ the situation is reversed: the
pressure, although higher than in the previous cases, is not able to oppose
sensibly the infall of BM and we end up with a distribution in the DM which is
more spread than that in the BM. This behaviour is confirmed by the value of
the bias which, for $n=-1$ is larger than one at all scales.

Our analysis is completely scale-free and the final distribution of the gas
temperature depends only on the assumed power spectrum. The highest values of
the temperature are obtained for the case $n=-1$ and they correspond to the
high density regions. As the spectral index increases, large-scale motion is
reduced and this decreases the height of the baryonic peaks and therefore also
the final heating of the gas is smaller. Dissipative processes present in the
collisional component allow the transformation of kinetic into internal energy
and this makes possible the formation of narrow baryonic peaks even in the
absence of cooling processes not included in the present work. In our
simulations pressure forces have never been able to equal gravitational forces
and so the motion of infall is expected to continue even in presence of high
temperatures. The inclusion of cooling terms would amplify this effect and
possibly cause a fragmentation of baryonic peaks. Work is presently in progress
for testing the effect of cooling and heating processes in the evolution of
one-dimensional cosmological perturbations. This, however, will require the
specification of the physical scales involved in the problem and therefore the
scale-free character of the present study will then be lost.

\section* {Acknowledgments} We thank Francesco Lucchin and Sabino Matarrese for
discussions. This work was partially supported by Italian MURST.

\bigskip


\section*{Figure captions}
\noindent
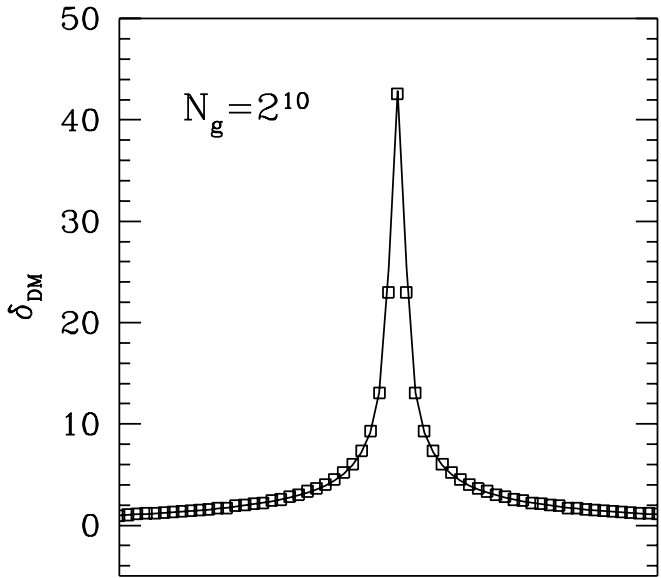
{\bf Figure 1.}
Sinusoidal test for the single dark matter component: distribution of density
fluctuations just before shell crossing. Comparison between the results
obtained from N-body (solid line) and Zel'dovich approximation (open squares),
when two different numbers of grid-points $N_g$ are adopted: $N_g=2^{10}$
(left) and $N_g=2^{15}$ (right).

\noindent
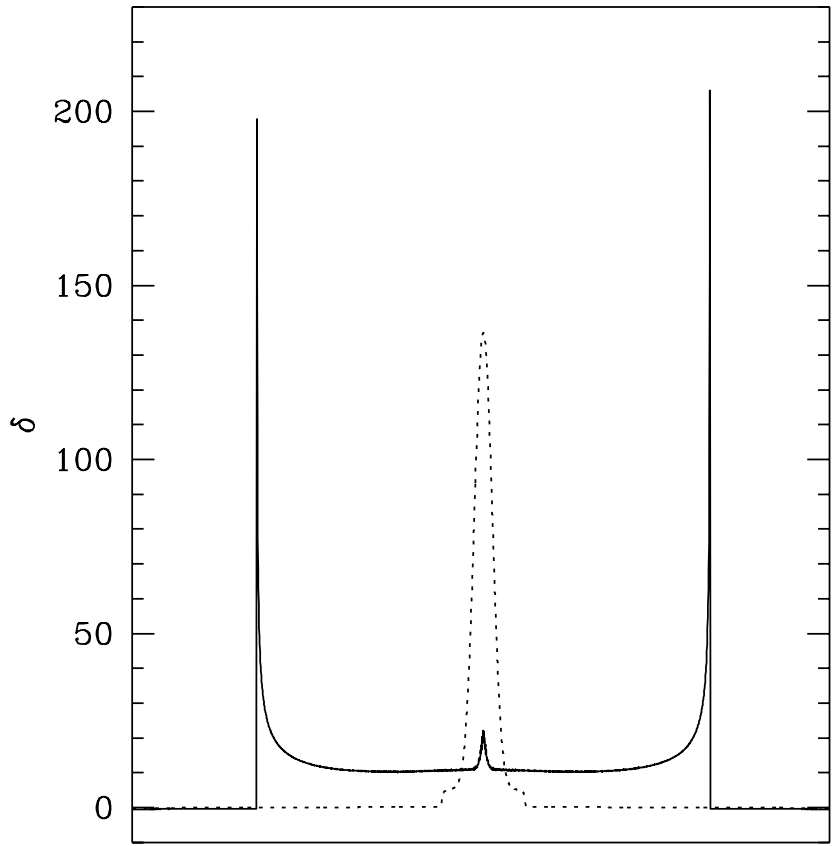
{\bf Figure 2.}
Sinusoidal test for the coupled evolution of baryonic and dark matter
components: distribution of density fluctuations after shell crossing. The
dotted and solid lines refer to the baryonic and dark matter components,
respectively.

\noindent
{\bf Figure 3a.}
The density fluctuation distribution at $a=0.5$ for the three different models:
$n=-1$ (top panel), $n=1$ (central panel) and $n=3$ (bottom panel). The dotted
and solid lines refer to the baryonic and dark matter components, respectively.
Only about a tenth of the whole grid is displayed.

\noindent
{\bf Figure 3b.}
As Figure 3a, but  at the final time $a=1$.

\noindent
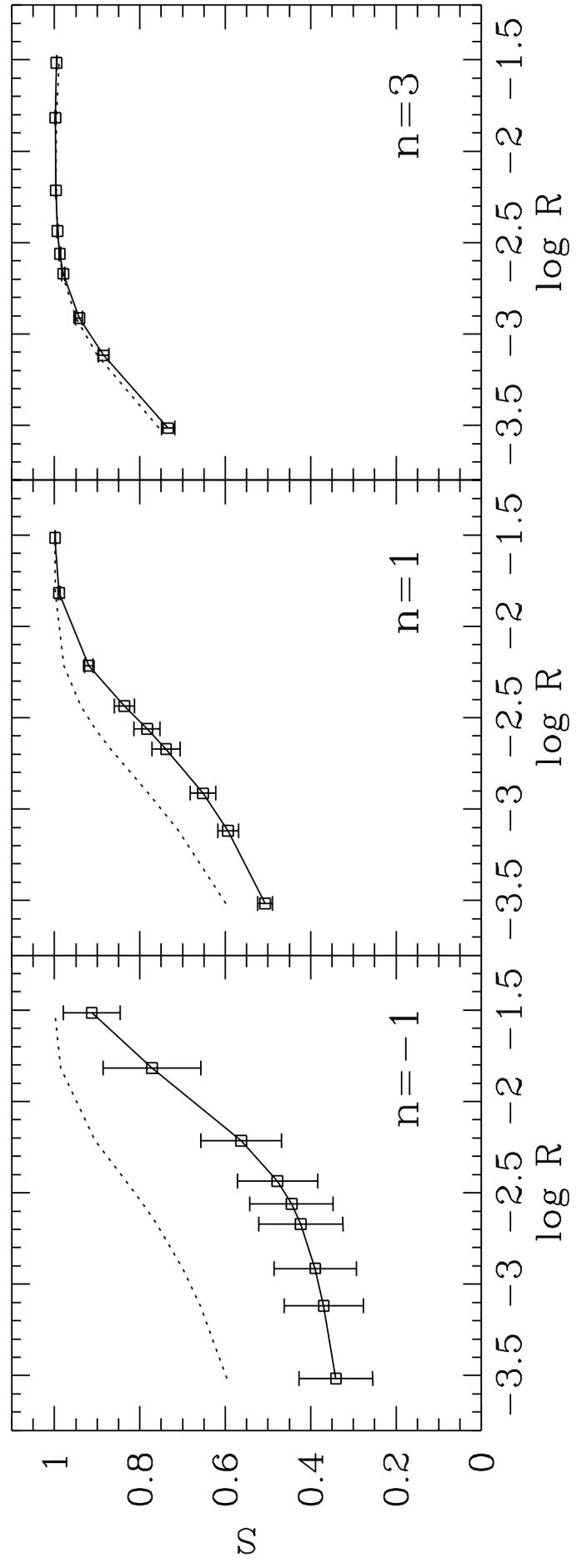
{\bf Figure 4.}
The cross-correlation coefficient $S$ as a function of the Gaussian filtering
radius $R$ (in units of the whole grid $L$) at the time $a=0.5$ (dotted line)
and $a=1$ (solid line)  for the different models: $n=-1$ (left), $n=1$ (centre)
and $n=3$ (right). Error bars (displayed for clarity only at $a=1$) represent
the r.m.s. scatter between the different simulations.

\noindent
{\bf Figure 5.}
The number of peaks $N_{pk}$ in dark (solid line) and baryonic (dotted line)
components as a function of their height $\delta$ when  $a=0.5$ (left column)
and $a=1$ (right column) for the different models: $n=-1$ (top), $n=1$ (centre)
and $n=3$ (down).

\noindent
{\bf Figure 6.}
The distribution of baryonic density fluctuation $\tilde \varrho_{_{BM}}$,
pressure $\tilde p$ and temperature $T$ for the case $n=-1$ at $a=1$. The whole
grid is displayed. Pressure is in arbitrary units while temperature is in
Kelvin degrees.

\noindent
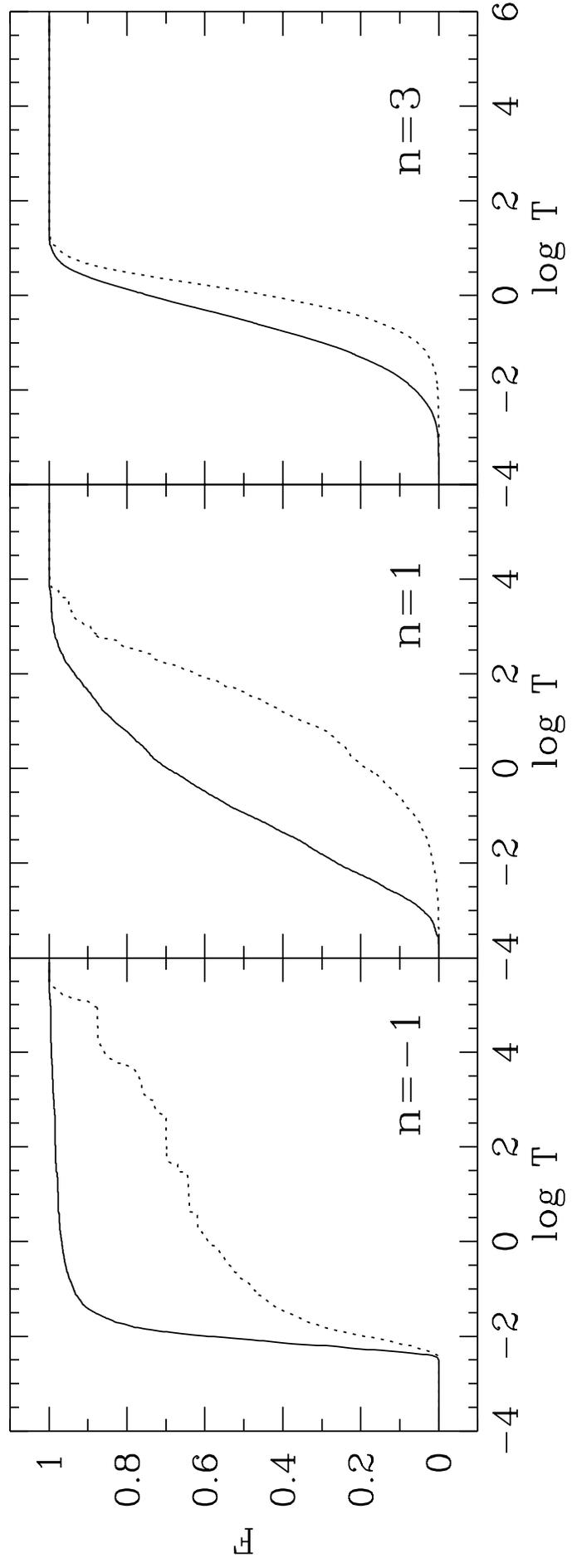
{\bf Figure 7.}
The cumulative volume filling factor (solid line) and mass fraction (dotted
line) as a function of the temperature $T$ at the final time $a=1$ for the
different models: $n=-1$ (left), $n=1$ (centre) and $n=3$ (right).

\noindent
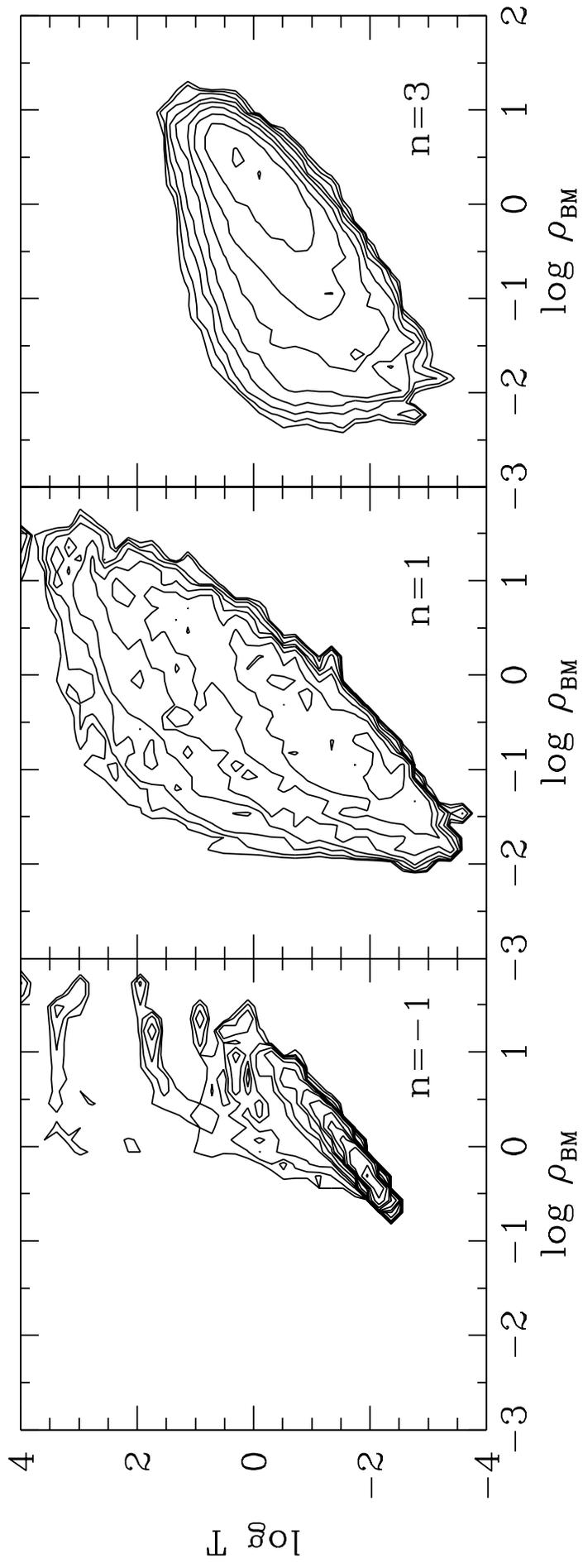
{\bf Figure 8.}
Contour plots for the number of grid-points characterized by a given
temperature $T$ and baryonic density $\tilde \varrho$ at the time $a=1$ for the
different models: $n=-1$ (left), $n=1$ (centre) and $n=3$ (right). The contour
levels are defined as follows: $10^{(I/4)}$, where $I$ is a positive integer;
the outermost contour corresponds to  $I=4$ and contours inside it have
gradually increasing $I$.

\noindent
{\bf Figure 9.}
The time evolution of the power-spectrum $P(k)$ for baryonic (left column)  and
dark matter (right column) component for the different models: $n=-1$ (top),
$n=1$ (centre) and $n=3$ (down). The curves refer to  different epochs:
$a=0.09, 0.17, 0.25, 0.33, 0.5$ (dotted lines from down to top) and $a=1$
(solid line).

\noindent
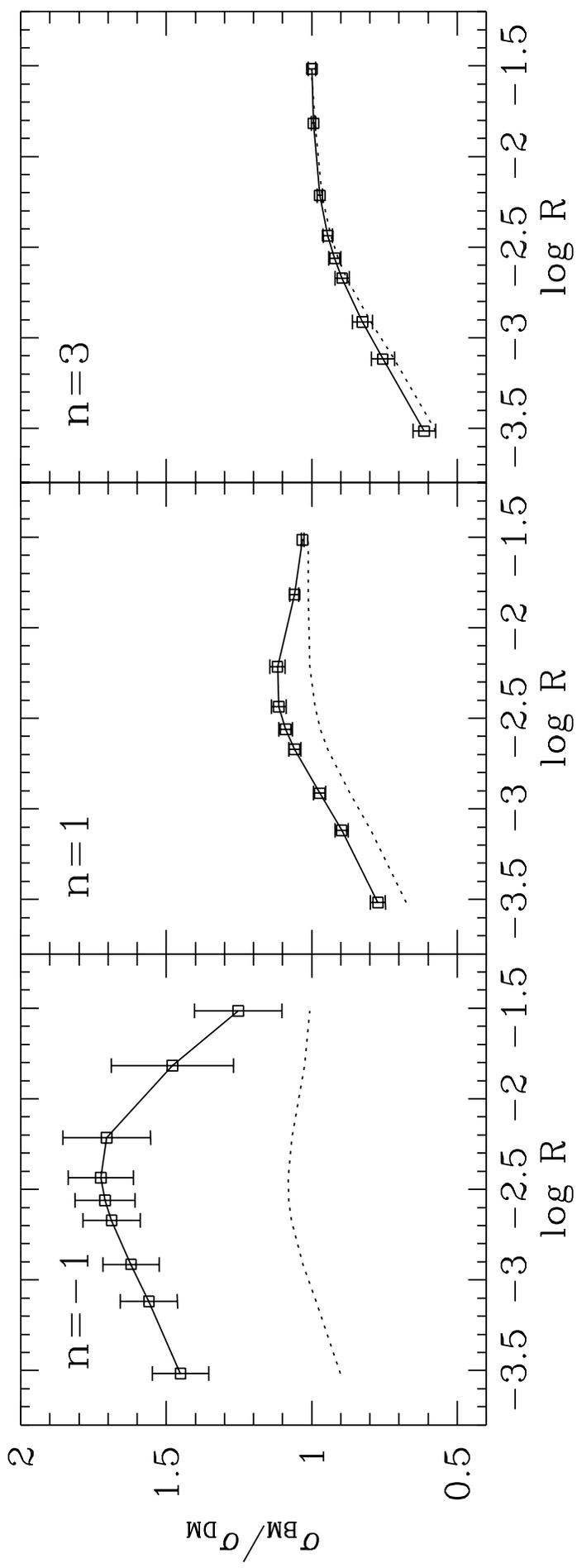
{\bf Figure 10.}
The relative bias $b\equiv \sigma_{_{BM}}/\sigma_{_{DM}}$ as a function of the
filtering scale $R$ at $a=0.5$ (dotted line) and $a=1$ (solid line) for the
different models: $n=-1$ (left), $n=1$ (centre) and $n=3$ (right). Error bars
(displayed for clarity only at $a=1$) represent the r.m.s. scatter between the
different simulations.

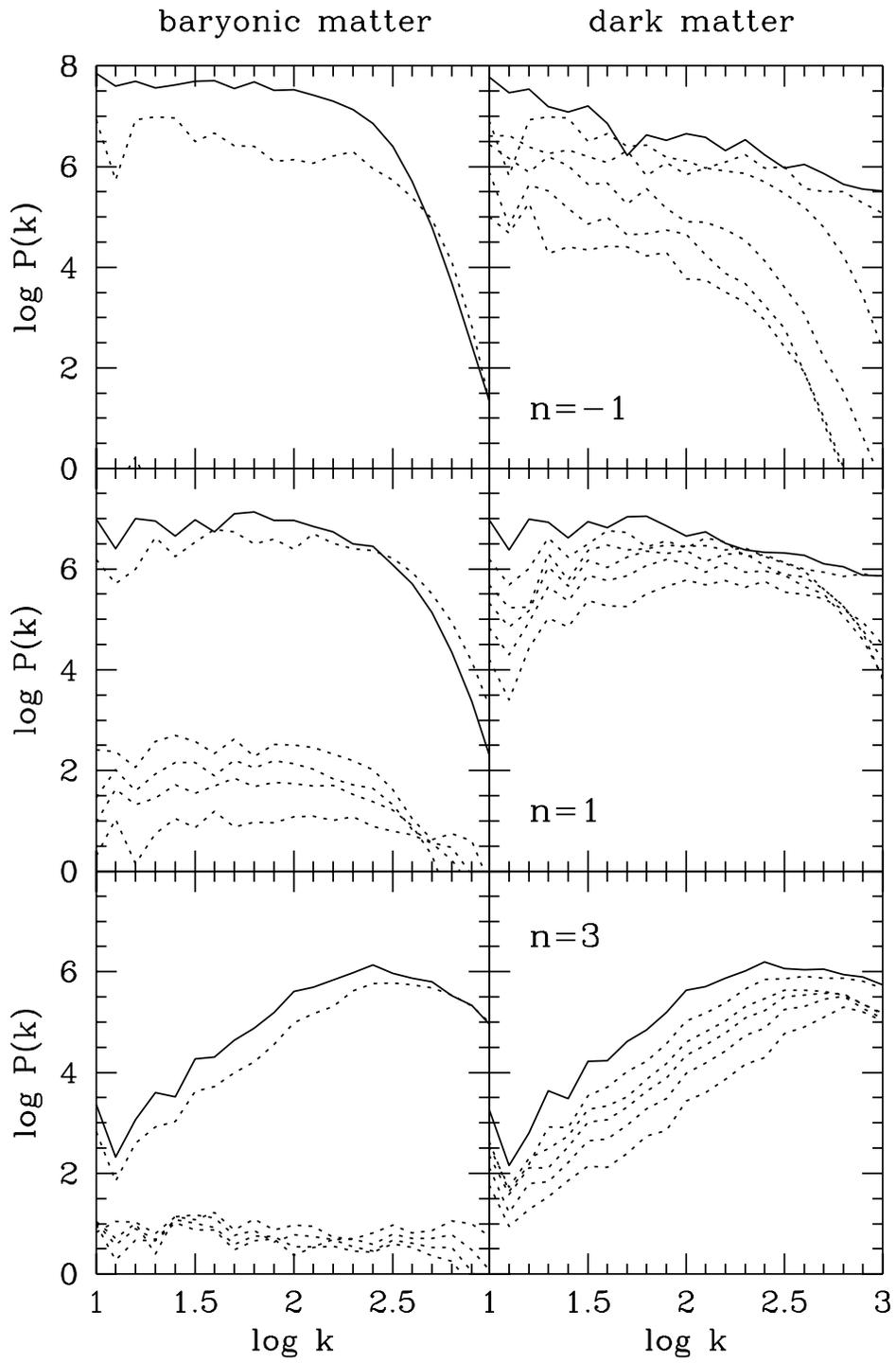
\end{document}